# Downlink Performance Enhancement of High Velocity Users in 5G Networks by Configuring Antenna System

Mariea Sharaf Anzum[1], Moontasir Rafique[2],

Md. Asif Ishrak Sarder[3], Fehima Tajrian[4], Abdullah Bin Shams [5]

[1234] Islamic University of Technology, Gazipur-1704, Bangladesh

[1]marieasharaf@iut-dhaka.edu, [2]moontasir@iut-dhaka.edu,

[3]asifishrak@iut-dhaka.edu, [4]fehimatajrian@iut-dhaka.edu

[5]University of Toronto, Toronto, ON M5S 3G8, Canada
[5]abdullahbinshams@gmail.com

**Abstract.** A limitation of bandwidth in the wireless network and the exponential rise in the high data rate requirement prompted the development of Massive Multiple-Input-Multiple-Output (MIMO) technique in 5G. Using this method, the ever-rising data rate can be met with the increment of the number of antennas. This comes at the price of energy consumption of higher amount, complex network setups and maintenance. Moreover, a high-velocity user experiences unpredictable fluctuations in the channel condition that deteriorates the downlink performance. Therefore, a proper number of antenna selection is of paramount importance. This issue has been addressed using different categories of algorithms but only for static users. In this study, we proffer to implement antenna diversity in closed loop spatial multiplexing MIMO transmission scheme by operating more number of reception antennas than the number of transmission antennas for ameliorating the downlink performance of high-velocity users in case of single user MIMO technology. In general, our results can be interpreted for large scale antenna systems like Massive MIMO even though a 4×4 MIMO system has been executed to carry out this study here. Additionally, it shows great prospects for solving practical-life problems like low data rate and call drops during handover to be experienced by cellular users traveling by high-speed transportation systems like Dhaka Metro Rail. The cell edge users are anticipated to get benefits from this method in case of SU-MIMO technology. The proposed method is expected to be easily implemented in the existing network structures with nominal difficulties.

# 1  INTRODUCTION

The ever-increasing cellular devices and exponential rise of wireless connection, demand for lower latency, higher spectral efficiency and ultra-high data speed. 5G technology is expected to meet these requirements. Thus, massive multiple-input-multiple-output (MIMO), an extension of MIMO technique becomes an essential requirement of this new standard technology. In the massive MIMO system, base stations use a huge number of antenna arrays to connect with users. These huge number of base station antennas can enable energy to concentrate in a small region bringing better improvements by several folds in transmission gain and user-throughput than a MIMO system. By integrating the single user MIMO (SU-MIMO) technique with the spatial multiplexing techniques, multiple number of non-identical data streams can be transmitted over various antennas to a single user equipment (UE). This results in a throughput gain with improved spectral efficiency.

However, practical implementation of a high number of antennas requires a larger amount of resources, high power supply, highly complex system and larger-scale antenna channel estimates. The increasing number of users is adding to these limitations due to loss in desired QoS (quality of service) resulting from spatial interference. For improving the QoS for each user, particle swarm optimization can be used to select the minimum number of transmitter antenna elements[1] . SUS algorithm and JASUS with a pre-coding scheme could be combined to select a limited number of antennas for specific users for lessening the complexity of the whole system delivering maximum average sum rate [2]. Then again, the large number of base station antennas contribute to the degradation of energy efficiency. This problem can be solved using the antenna selection method based on channel state information [3]. Additionally, a deep learning strategy has been proposed to optimize antenna selection pattern for massive MIMO channel extrapolation [4]. The aforementioned research works proffered optimum antenna selection patterns or algorithms for mitigating some of the limitations of Massive MIMO mentioned earlier, but did not consider practical scenarios like UE mobility which was investigated in recent research works. The high velocity users experience a degradation of performance of the schedulers, spatial multiplexing techniques and network capacity because mobility of UE causes Doppler shift which results in rapid variations in channel quality [5]-[6]. Due to the frequent and rapid variations of the channel quality, the optimum antenna combination also changes over the multiple transmission time interval's (TTIs). Therefore, the optimum antenna combinations may not be feasible for practical implementation without considering mobility. Moreover, deep learning algorithms can cause high latency in real-time application which contradicts one of the goals of 5G technology.

In this paper, we have proposed to implement antenna diversity in spatial multiplexing technique by keeping the number of receiver antennas more than the number of transmitter antennas. This diversity of antennas has enabled overall throughput for high-velocity users to improve. This simple strategy will work under any scheduler and transmission schemes considering users with low, medium, and high velocity. To imitate a system with a high number of transmission and reception antennas and to circumvent the simulation complexity, a 4×4 MIMO system has been implemented to conduct our study. In general, our results can be interpreted for large scale antenna systems like Massive MIMO. It is believed that the proposed strategy will contribute to solving real-life problems like call drops, low data speed to be faced by cellular users traveling through high-speed transportation systems like Dhaka Metro Rail.

The remaining parts of the proffered paper are assembled as the following sequence. Section 2 contains the system model where spatial multiplexing, antenna diversity, resource scheduling schemes, network model and various performance parameters are discussed. After that, simulation model is discussed in section 3 before discussing the simulation results in section 4. In the end, conclusions are presented in section 5.

## 2 SYSTEM MODEL

### 2.1 Spatial Multiplexing

Spatial multiplexing is one of the transmission modes used in a communication system. Generally, the data which needs to be sent to any UE are divided into several streams and sent over multiple channels using same frequency band.
Through spatial multiplexing, the ability to use several channels simultaneously for transmitting data helps to increase system capacity. Based on theoretical deduction, the capacity of any channel increases with the increment of number of data streams keeping a linear relationship according to the following equation[7]:

$$C = MB \log_2 \left(1 + \frac{S}{N}\right) \qquad (1)$$

Here, capacity is represented by $C$ for bandwidth $B$ whereas number of data streams is $M$. And $\frac{S}{N}$ is signal to noise ratio.

There are mainly two categories of spatial multiplexing-close loop spatial multiplexing (CLSM) and open loop spatial multiplexing (OLSM). For the proffered strategy, CLSM is adopted as the spatial multiplexing technique.

***Close loop spatial multiplexing (CLSM)***: The maximum number of data streams $M$ in $T$ transmitter and $R$ receiver network ($T \times R$) that can be transmitted over the network follows the given condition in case of CLSM.[5].

$$M \leq min(T, R) \qquad (2)$$

Under CLSM, UE reported CQI gives indication about channel condition which is utilized to select the suitable MCS (Modulation and Coding Scheme) by base station for users. Then rank indicator is utilized to select the number of layers under the selected MCS and channel. In this paper, MCS level can be selected as 16-QAM for the simulation. And then PMI feedback helps UE to adjust quickly to the frequently changing condition of the channel.

### 2.2 Network Model

A typical network comprised of 19 macro cell base stations is used here. Base stations are arranged in a hexagonal manner creating a two-tier network. The distance between any two base stations is 500 m. Every base station antenna is taken as a tri-sector antenna to achieve 360-degree coverage. An architecture of the adopted model has been depicted in **Fig. 1**. To design these antennas, the Kathrein 742215 antenna model is utilized here[8].

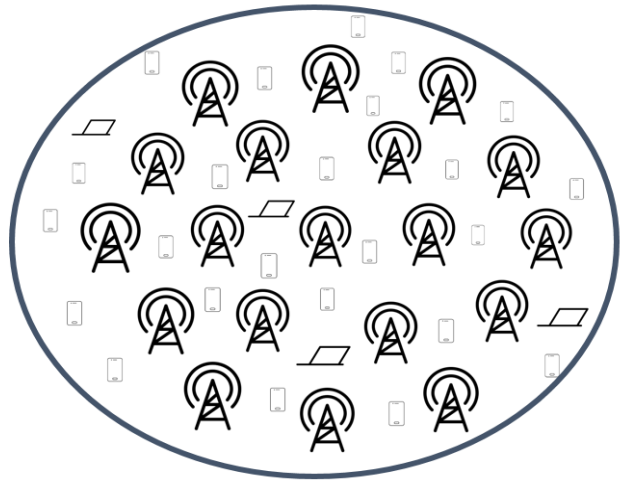

**Fig. 1.** Network Architecture

### 2.3 Resource Scheduling

Resource scheduling method is basically utilized to allocate resources among the users in any cell in a specific way considering some criteria like system efficiency, users demand etc. Round Robin (RR) and Proportional fair (PF) have been adopted for our simulation results.

*Proportional Fair & Round Robin*

The main consideration of RR is to ensure optimum fairness among all users in a cell adapting the cyclic distribution of resources. On the other hand, proportional fair primarily considers to boost throughput of users keeping fairness as the secondary consideration. Each user equipment is taken as the priority coefficient from the priority function given in the following equation.[6]

$$P = \frac{T^\alpha}{R^\beta} \tag{3}$$

Here, $T$ is feasible throughput and $R$ is average data rate. Using parameters, $\alpha$ and $\beta$ fairness can be tuned. In case of PF, $\alpha \approx 1$ and $\beta \approx 1$ are used whereas $\alpha \approx 0$ and $\beta \approx 1$ are used in case of RR technique[5]. But in order to enhance user throughput and fairness of resource allocation, an improved PF scheduling algorithm has been proffered where average user throughput $T_k(t)$ can be calculated using the following equation [9].

$$T_k = \left(1 - \frac{1}{t_c}\right) T_k(t) + \frac{1}{t} \sum_{s=1}^{s} R_{s,k}(t) \tag{4}$$

Here, the throughput of the user $k$ at sub band $s$ is represented by $R_{s,k}(t)$, throughput averaging time window is defined by $t_c$. The value of $t_c$ can be selected to carry out an optimum trade-off between fairness and system capacity. On the other hand, RR distributes resources among users without considering channel condition in a cyclic manner. As a result, it can ensure fairness amidst users but degrades user throughput.

### 2.4 Key Performance Parameters

*Average UE Throughput:* The position of users in the whole network is random. So, the distance between them and the base stations is not fixed which results in variations in path loss. Consequently, SINR and throughput of a UE have a huge range of different values. Moreover, it is well known that throughput of a T×R system is proportional to min(T, R) where T is transmitter antenna number and R is receiver antenna number. Then again, average throughput of UE ($T_{AVG}$) can be expressed like the following equation[10].

$$T_{AVG} = \frac{\sum_{k=1}^{n} T_k}{n} \tag{5}$$

Here, total throughput of $k^{th}$ user is represented by $T_k$ and the total number of users is $n$.

*Cell edge throughput:* Users near the edge of a serving cell feel signals from the neighboring cells as interferences for them. Adding to that, the strength of the signal degrades owing to the distance from the base station causing lower speed of data. Cell edge throughput is considered as the five percent of the throughput ECDF of UE. For avoiding call drop, continuous coverage and minimum data rate are required during handover.

*Spectral Efficiency:* Spectral efficiency can be stated as the speed of data over a specific bandwidth. This parameter can be represented by the following equation[7].

$$s = \frac{\sum_{k=1}^{n} T_k}{BW} \qquad (6)$$

Here, system bandwidth is $BW$ and total throughput for $k^{\text{th}}$ user is $T_k$. Thus, when total throughput over a specific bandwidth increases, spectral efficiency increases.

*Fairness Index:* Fairness index is the parameter which is utilized to dictate how the resources may be divided among the UEs. A well-known method, Jain's Fairness Index is adopted to examine the fairness in terms of resources among the UEs. Jain's Fairness Index can be calculated for $n$ users from the following equation[11].

$$J(T) = \frac{[\sum_{k=1}^{n} T_k]^2}{n[\sum_{k=1}^{n} T_k^2]} \qquad (7)$$

Here, average throughput is represented by $T_k$ for $k^{\text{th}}$ user. $J(T) = 1$ if all the UEs can achieve the same share of resources distributed among them.

## 3      SIMULATION MODEL

At first, performance parameters are numerically investigated under proportional fair resource scheduling technique for 2×2, 2×3, 2×4, 4×2, 4×3, 4×4 MIMO schemes. After that, the investigation was done for the MIMO schemes for round robin resource scheduling technique at second phase. In both of the phases, impact of UE mobility spanning over a range of 0-120 kmph velocity was observed and then results were plotted. There were 10 UEs per sector with a total of 570 UEs arbitrarily taken within the geometrical area of the network during the simulation. To integrate mobility of UEs in the simulations, random walk model has been selected. Using random walk model, a user can be assumed to take a random step away from the previous position in each period. For link prediction for UEs, Mutual Information Effective SINR Mapping (MIESM) is chosen here owing to its better accuracy than other ESM algorithms especially for higher modulation schemes[12]. Simulations for round robin have been done for 20 TTIs whereas

50 TTIs have been considered for proportional fair. The Vienna LTE-Advanced simulator has been utilized here[13]. The macroscopic path loss model for the macro-cells considering an urban environment can be represented by the following equation [7].

$$L = 40(1 - 4 \times 10^{-3} h_{Bs}) \log_{10}(R) - 18 \log_{10}(h_{Bs}) + 21 \log_{10}(f_c) + 80 dB \quad (8)$$

Here, $R$ is taken as the separation between the UE and the base station in kilometers, carrier frequency in MHz is represented by $f_c$ and height of the antenna is $h_{Bs}$ in meters. Rest of the parameters are displayed in **Table 1.**

Table 1. Parameters for the simulation of the network

| Simulation Parameters | |
|---|---|
| **Channel Model** | WINNER+ |
| **Frequency** | 2.45 GHz |
| **Bandwidth** | 20MHz |
| **No. of transmitter/receiver** | 4 |
| **Simulation Time** | 50TTI (PF), 50TTI (RR) |
| **BS height** | 20 m |
| **Transmission mode** | CLSM(2 × 2, 2 × 3, 2 × 4, 4 × 2, 4 × 3, 4 × 4) |
| **BS power** | 45 dB |
| **Receiver height** | 1.5 m |
| **Antenna azimuth offset** | 30 degree |
| **BS transmitter power** | 45dBm |
| **Antenna Gain** | 15dBi |

## 4 RESULTS AND DISCUSSIONS

Impact of mobility on average UE throughput under PF and RR with different MIMO schemes can be observed from **Fig. 2(a)** and **2(b)** respectively. Due to mobility, variations in channel condition cause SINR to have low values. Thus, average throughput degrades at increasing velocity under both schedulers. From **Fig. 2(a),** it can be observed that PF achieves better average UE throughput for the whole velocity range of 0-120kmph under 2×4, 2×3 and, 2×2 MIMO schemes compared to 4×2, 4×3 and, 4×4 MIMO schemes. On the other hand, RR shows best performance for 4×4 MIMO at low velocity but seems to have degrading performance compared to 2×4 and 2×3 MIMO at high velocity as displayed in **Fig. 2(b).** Because more number of receiving antennas compared to number of transmitting antennas enable a receiver to have more options to choose signals with better SINR. Consequently, a receiver can have enhanced reception quality, better link performance along with better throughput.

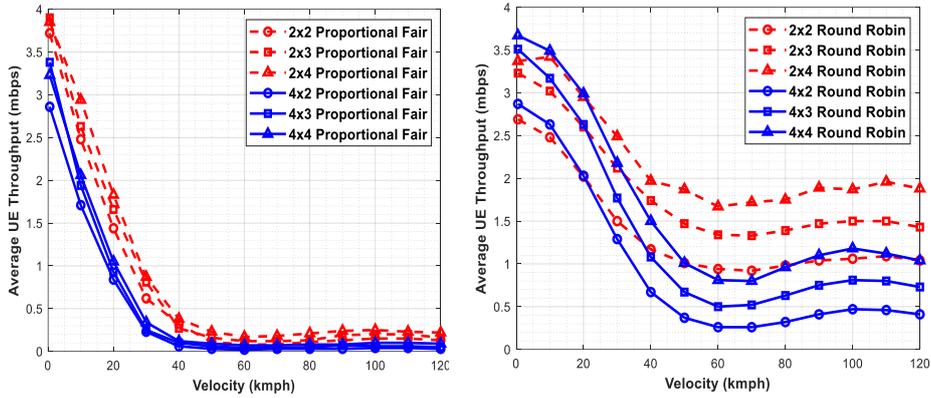

**Fig. 2.** Impact of mobility on Average UE Throughput under **(a)** PF and **(b)** RR scheme

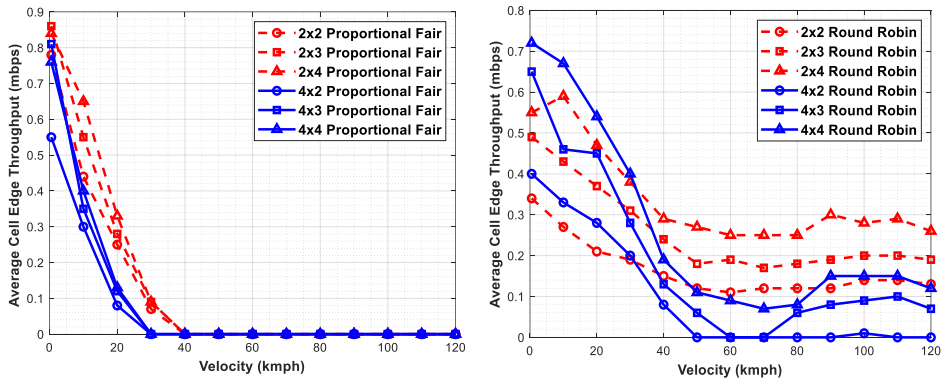

**Fig. 3**. Impact of mobility on cell edge throughput under **(a)** PF and **(b)** RR scheme

PF shows good cell edge throughput at low velocity but ends up going to zero cell edge throughput at increasing velocity under all MIMO schemes as shown in **Fig. 3(a)**. The same does not happen in case of RR as it does not consider channel condition. From **Fig. 3(b)**, it can be observed that RR shows a slow decrement and ends up with very low but necessary to pursue handover from one cell to another at the cell edge. Moreover, it can be noticed that cell edge throughput shows slow a decline under 2×4 and 2×3 MIMO and holds up better performance than other MIMO schemes as velocity increases up to 120kmph. This occurrence indicates that increasing the number of receiver antennas than that of transmitter antennas is an efficient way to reduce call drops and link failures due to better cell edge throughput ensuring a ubiquitous network.

Then the effect of mobility on spectral efficiency can be observed from **Fig. 4(a)** and **Fig. 4(b)** under both scheduling techniques along with different antenna configurations. Spectral efficiency seems to decline a lot as velocity of a user increases under PF scheduling technique. Here, 2×4 MIMO seems to show better performance than any other schemes under PF over the whole velocity range of 0 to 120kmph.From before, it was noted that average throughput was also better for antenna configuration where transmitting antennas are less than receiving antennas. Consequently, spectral efficiency shows the same characteristic for PF. Then coming to RR, it can be observed that 4×2 and 4×3 MIMO scheme seem to show better performance at low velocity due to good SINR and throughput but seems to show degrading performance than 2×4 and 2×3 at high velocity. This incidence happens because receiver gets to avoid the signals with low SINR value and select the strong signal with robust link performance when receiving antennas are more in number. As a result, spectral efficiency is better whenever receiver antennas are more than transmitter antennas as high UE throughput and cell edge throughput can be achieved efficiently utilizing that specific bandwidth.

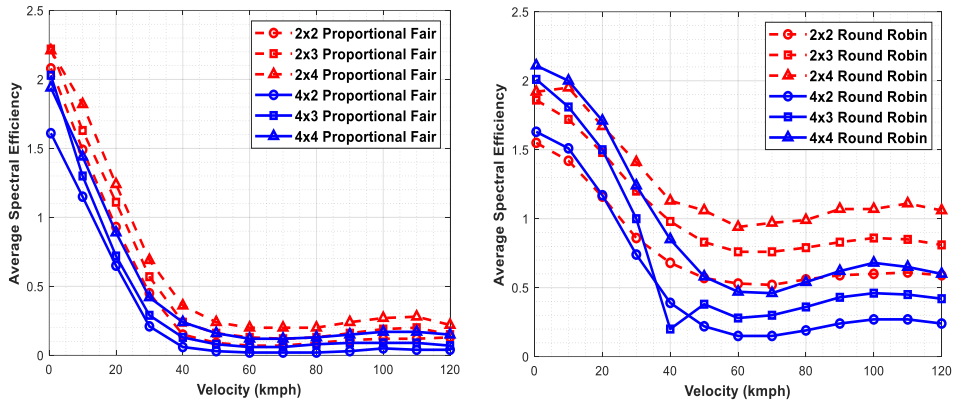

**Fig. 4.** Impact of mobility on average spectral efficiency under **(a)** PF and **(b)** RR scheme

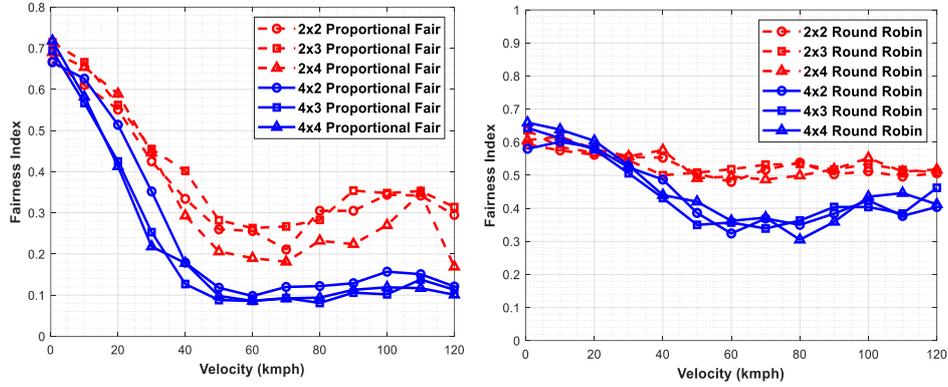

**Fig.5.** Impact of mobility on fairness index under **(a)** PF and **(b)** RR scheme

Finally, from **Fig. 5. (a)** and **(b)**, performance of different antenna configurations can be analyzed in terms of fairness index under both schedulers. Under RR, fairness index does not drop that much comparing to PF as RR does not take channel condition into consideration. Moreover, both schedulers seem to achieve better fairness index whenever transmitting antennas are less than receiving antennas over the investigated range of velocity due to better average throughput and cell edge throughput. Additionally, more receiving antennas can help achieve better SINR which ultimately results in UE sending better CQI value to the base station which allocates the resources accordingly.

**Table 2.** Comparison with previous research works

| Research works | Transmission mode | Number of transmitting and receiving antennas | Diversity in spatial multiplexing |
|---|---|---|---|
| [5] | CLSM and OLSM | 2×2 | No |
| [6] | Transmit diversity | 2×2 | No |
| [7] | CLSM | 4×4 | No |
| **This paper** | **CLSM** | **2×2, 2×3, 2×4, 4×2, 4×3, 4×4** | **Yes** |

**Table 2** mentions some of the previous works where impact of mobility has been considered to analyze performance of downlink of cellular networks. Previous works mentioned in the table were proposed for LTE and LTE Advanced networks whereas this work is proposed for upgrading downlink performance of 5G network. None of these works investigated different antenna configurations in high velocity for single user Massive MIMO system. It is best to our knowledge that antenna diversity in SU-MIMO technology in Massive MIMO setup has not ever been investigated with respect to velocities ranging from 0-120 km/h under different resource scheduling schemes. Additionally, evaluating all performance parameters brought up above, it can be suggested that implementing antenna configuration where receiving antennas are more than transmitting antennas ensures enhanced downlink performance for a high-velocity user reducing multipath fading and channel fluctuations. It is also believed that introducing the concept of antenna diversity in SU-MIMO technology will pave the way for a cheaper and simple solution for high velocity users giving better possibility of reception quality at the receiver end through both of the resource scheduling schemes considered in this study.

## 5    CONCLUSION

In this paper, we proffer to introduce antenna diversity in spatial multiplexing MIMO transmission scheme by operating more number of reception antenna than the number of transmission antenna to improve the downlink performance of high-velocity users. To circumvent the simulation complexity, a 4×4 MIMO system has been implemented to conduct the study. The simulation results show that this technique can work independent of the type of applied resource scheduler and the same is expected under any transmission scheme. A linear increase in data rate with higher reception antennas is also observed. The proposed method can be easily implemented in the existing network architectures with minimal difficulties. Also, it has the potential for solving real-life problems like call drops and low data rate to be experienced by cellular users traveling through high-speed transportation systems like Dhaka Metro Rail.

**References**

[1]    D. Park, "Transmit antenna selection in massive MIMO systems," *Int. Conf. Inf. Commun. Technol. Converg. ICT Converg. Technol. Lead. Fourth Ind. Revolution, ICTC 2017*, vol. 2017-Decem, no. September, pp. 542–544, 2017, doi: 10.1109/ICTC.2017.8191036.

[2]    T. A. Sheikh, J. Bora, and M. A. Hussain, "Combined User and Antenna Selection in Massive MIMO Using Precoding Technique," *Int. J. Sensors, Wirel. Commun. Control*, vol. 9, no. 2, pp. 214–223, 2018, doi: 10.2174/2210327908666181112144939.

[3]    G. Jin, C. Zhao, Z. Fan, and J. Jin, "Antenna Selection in TDD Massive MIMO Systems," *Mob. Networks Appl.*, 2019, doi: 10.1007/s11036-019-01297-5.

[4]    Y. Yang, S. Zhang, F. F. Gao, C. Xu, J. Ma, and O. A. Dobre, "Deep Learning Based Antenna Selection for Channel Extrapolation in FDD Massive MIMO," *12th Int. Conf. Wirel. Commun. Signal Process. WCSP 2020*, pp. 182–187, 2020, doi:

10.1109/WCSP49889.2020.9299795.

[5] A. Bin Shams, S. R. Abied, M. Asaduzzaman, and M. F. Hossain, "Mobility effect on the downlink performance of spatial multiplexing techniques under different scheduling algorithms in heterogeneous network," *ECCE 2017 - Int. Conf. Electr. Comput. Commun. Eng.*, no. February, pp. 905–910, 2017, doi: 10.1109/ECACE.2017.7913032.

[6] A. Bin Shams, S. R. Abied, and M. A. Hoque, "Impact of user mobility on the performance of downlink resource scheduling in Heterogeneous LTE cellular networks," *2016 3rd Int. Conf. Electr. Eng. Inf. Commun. Technol. iCEEiCT 2016*, no. September, 2017, doi: 10.1109/CEEICT.2016.7873091.

[7] A. Bin Shams, M. R. Meghla, M. Asaduzzaman, and M. F. Hossain, "Performance of coordinated scheduling in downlink LTE-a under user mobility," *4th Int. Conf. Electr. Eng. Inf. Commun. Technol. iCEEiCT 2018*, no. September, pp. 215–220, 2019, doi: 10.1109/CEEICT.2018.8628126.

[8] F. Gunnarsson *et al.*, "Downtilted base station antennas - A simulation model proposal and impact on HSPA and LTE performance," *IEEE Veh. Technol. Conf.*, pp. 1–5, 2008, doi: 10.1109/VETECF.2008.49.

[9] M. R. Hojeij, C. Abdel Nour, J. Farah, and C. Douillard, "Weighted Proportional Fair Scheduling for Downlink Nonorthogonal Multiple Access," *Wirel. Commun. Mob. Comput.*, vol. 2018, 2018, doi: 10.1155/2018/5642765.

[10] A. Noliya and S. Kumar, *Performance Analysis of Resource Scheduling Techniques in Homogeneous and Heterogeneous Small Cell LTE-A Networks*, vol. 112, no. 4. Springer US, 2020.

[11] R. Jain, D. Chiu, and W. Hawe, "A Quantitative Measure Of Fairness And Discrimination For Resource Allocation In Shared Computer Systems." 1998, [Online]. Available: http://arxiv.org/abs/cs/9809099.

[12] F. L. Aguilar, G. R. Cidre, J. M. L. López, and J. R. Paris, "Mutual Information Effective SNR Mapping algorithm for fast link adaptation model in 802.16e," *Lect. Notes Inst. Comput. Sci. Soc. Telecommun. Eng.*, vol. 45 LNICST, no. January, pp. 356–367, 2010, doi: 10.1007/978-3-642-16644-0_31.

[13] M. Rupp, S. Schwarz, and M. Taranetz, *Signals and Communication Technology The Vienna LTE-Advanced Simulators Up and Downlink, Link and System Level Simulation*. .